\documentclass[10pt]{iopart}
\usepackage{epsfig,cite}
\newcommand{\ewxy}[2]{\setlength{\epsfxsize}{#2}\epsfbox[30 30 640 640]{#1}}
\newcommand{\beq}{\begin{equation}}
\newcommand{\eeq}{\end{equation}}
\newcommand{\ds}{\displaystyle}
\newcommand{\beqar}{\begin{eqnarray}}
\newcommand{\eeqar}{\end{eqnarray}}

\begin{document}

\title{ Anisotropic flow of strange particles at RHIC }
\author{
E~E~Zabrodin\dag\ddag, L~V~Bravina\dag\ddag, G~Burau\S, 
L~Bleibel\S, C~Fuchs\S, Amand~Faessler\S, 
}
\address{\dag\
         Department of Physics, University of Oslo, Oslo, Norway}
\address{\ddag\
         Institute for Nuclear Physics, Moscow State University, 
         Moscow, Russia}
\address{\S\
         Institut f\"ur Theoretische Physik, Universit\"at
         T\"ubingen, T\"ubingen, Germany}

\begin{abstract}
Space-time picture of the anisotropic flow evolution in Au+Au
collisions at BNL RHIC is studied for strange hadrons within the
microscopic quark-gluon string model. The directed flow of both mesons
and hyperons demonstrates wiggle structure with the universal antiflow 
slope at $|y| \leq 2$ for minimum bias events. 
This effect increases as the reaction becomes more peripheral.
The development of both components of the anisotropic flow is closely
related to particle freeze-out. Hadrons are emitted continuously, and
different hadronic species are decoupled from the system at different
times. These hadrons contribute differently to the formation and
evolution of the elliptic flow, which can be decomposed onto three
components: (i) flow created by hadrons emitted from the surface at
the onset of the collision; (ii) flow produced by jets; (iii)
hydrodynamic flow. Due to these features, the general trend in 
elliptic flow formation is that the earlier mesons
are frozen, the weaker their elliptic flow. In contrast, baryons
frozen at the end of the system evolution have stronger $v_2$.
\end{abstract}

\section{Introduction}
\label{sec1}
The transverse collective flow of particles is usually decomposed
onto isotropic radial flow and anisotropic components, such as 
directed flow, elliptic flow, etc.
Directed and elliptic flows are defined as the first and the second
harmonic coefficients, $v_1$ and $v_2$, of an azimuthal Fourier 
expansion of the particle invariant distribution \cite{VZ96}
\beq
\ds
E \frac{d^3 N}{d^3 p} = \frac{1}{\pi} \frac{d^2 N}{dp_t^2 dy} \left[
1 + 2 v_1 \cos(\phi) + 2 v_2 \cos(2 \phi) + \ldots \right] \ ,
\label{eq1}
\eeq
where $\phi$ is the azimuthal angle between the transverse momentum of
the particle and the reaction plane, and $p_t$ and $y$ is the
transverse momentum and the rapidity, respectively. The directed and
elliptic flows can be presented as
\beq \ds
v_1 \equiv \langle \cos{\phi} \rangle =
\left \langle \frac{p_x}{p_t} \right \rangle \ , \ \ 
v_2 \equiv \langle \cos{2 \phi} \rangle =
\left \langle \frac{p_x^2 -p_y^2}{p_t^2} \right \rangle \ .
\label{eq2}
\eeq
with $x$ being the impact parameter axis. The transverse momentum of 
a particle is simply $p_t = \sqrt{p_x^2 + p_y^2}$.

The overlapping area of two nuclei colliding with non-zero impact
parameter $b$ has a characteristic almond shape in the transverse 
plane. The fireball tries to restore spherical shape, provided the 
thermalization sets in rapidly and the hydrodynamic description is 
appropriate \cite{QM04}. When it becomes spherical, apparently, the 
elliptic flow stops to develop.  Therefore, $v_2$ can carry 
important information about the earlier phase of ultrarelativistic 
heavy-ion collisions, equation of state (EOS) of hot and dense 
partonic matter, and is expected to be a useful tool to probe the 
formation and hadronization of the quark-gluon plasma (QGP).

\section{Directed and Elliptic Flow}
\label{sec2}
\begin{figure}[htb]
\begin{minipage}[t]{65mm}
\ewxy{zab_fig1.epsi}{70mm}  
\vspace{-1.5cm}
\caption{
Directed flow $v_1(y)$ of $\phi, N, K$ in minimum bias Au+Au 
events at $\sqrt{s} = 130$ AGeV.
}
\label{fig1}
\end{minipage}
\hspace{\fill}
\begin{minipage}[t]{65mm}
\vspace{-8.0cm}
\ewxy{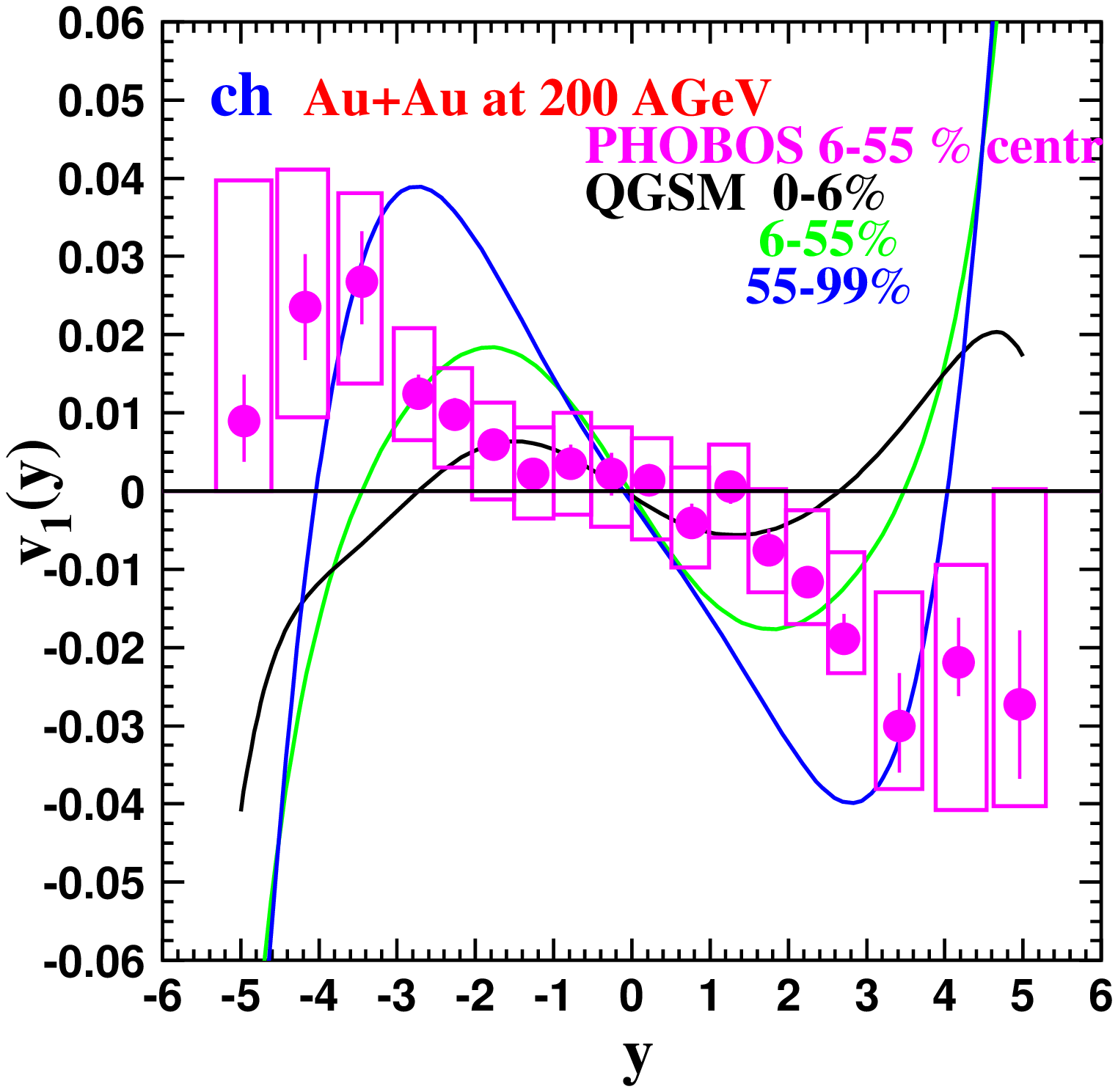}{67mm}  
\caption{
Centrality dependence of the $v_1^{ch}(\eta)$ in
Au+Au collisions at $\sqrt{s} = 200$ AGeV.
}
\label{fig2}
\end{minipage}
\end{figure}

Figure~\ref{fig1} shows the rapidity dependence of the directed flow
of $\phi, N$, and $K$ in minimum bias Au+Au collisions at $\sqrt{s} =
130$ AGeV. The slopes of the all distributions are
negative at $|y| \leq 2$, i.e. the antiflow component of the $v_1$
dominates over its normal counterpart (see \cite{dir_flow}). Similar
antiflow slopes of $v_1(y)$ are developed by $\pi$ and $\Lambda$
\cite{dir_flow,kflow}; its origin is traced to nuclear shadowing. At
midrapidity $|y| \leq 0.5$ the directed flow of all hadrons is quite 
weak. Figure \ref{fig2} depicts the simulation results for the 
$v_1(\eta)$ of charged hadrons compared to the experimental data from 
the PHOBOS \cite{phob_dir} Collaboration for 6\% to 55\% central
Au+Au collisions at $\sqrt{s} = 200$ AGeV
One can see that the model reproduces the $v_1$ data quite well both
qualitatively and quantitatively, although the maxima of the directed
flow around $|\eta| \approx 2$ are shifted to lower
pseudorapidities compared to the experimental data.
Similar antiflow alignment can be obtained also within the multi
module model (MMM) \cite{MCS02}, which is based on fluid dynamics
coupled to formation of colour ropes.
Microscopic models based on FRITIOF routine, e.g. UrQMD and AMPT, 
show a very flat and essentially zero directed flow \cite{BlSt02,ampt} 
in a broad range $|\eta| \leq 2.5$.
Although the data seem to indicate
antiflow behaviour for the directed flow of charged particles with
the possible flatness at $|\eta| \leq 1.5$, the measured signal is
quite weak, -- the magnitude of the flow is less than 1\% at $|\eta|
\leq 2$. Therefore, relatively large systematic error bars do not
permit us to disentangle between the different models.
The other features which should be mentioned here are
broadening of the antiflow region and increase of its strength as the
reaction becomes more peripheral.

\begin{figure}[htb]
\begin{minipage}[t]{65mm}
\hspace{1.0cm}
\ewxy{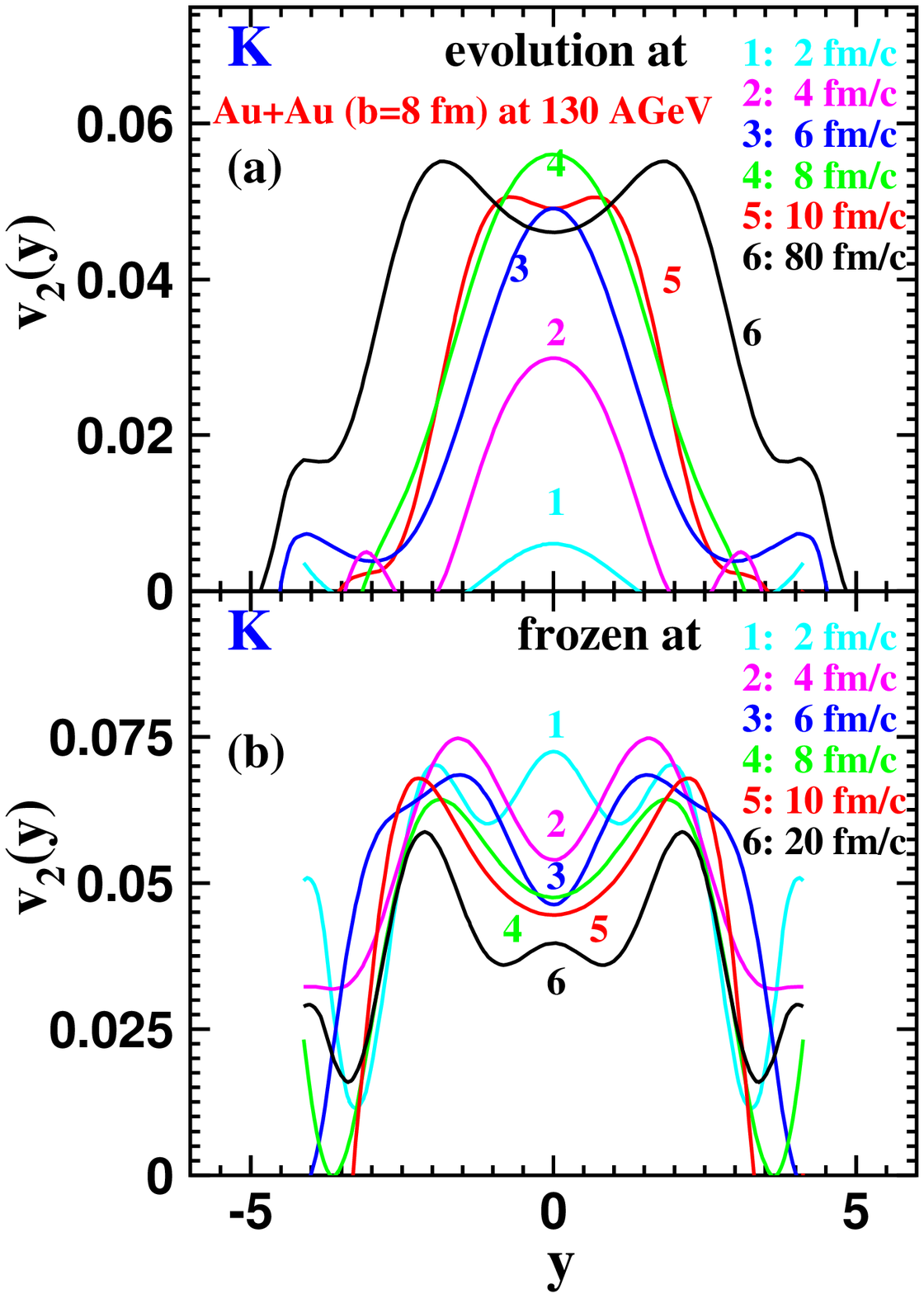}{72mm}  
\caption{
$d^2N/dx dy$ distributions for $\Lambda$ and $K$ in Au+Au
collisions with $b = 8$ fm at $\sqrt{s} = 130$ AGeV.
}
\label{fig3}
\end{minipage}
\hspace{\fill}
\begin{minipage}[t]{65mm}
\hspace{1.0cm}
\ewxy{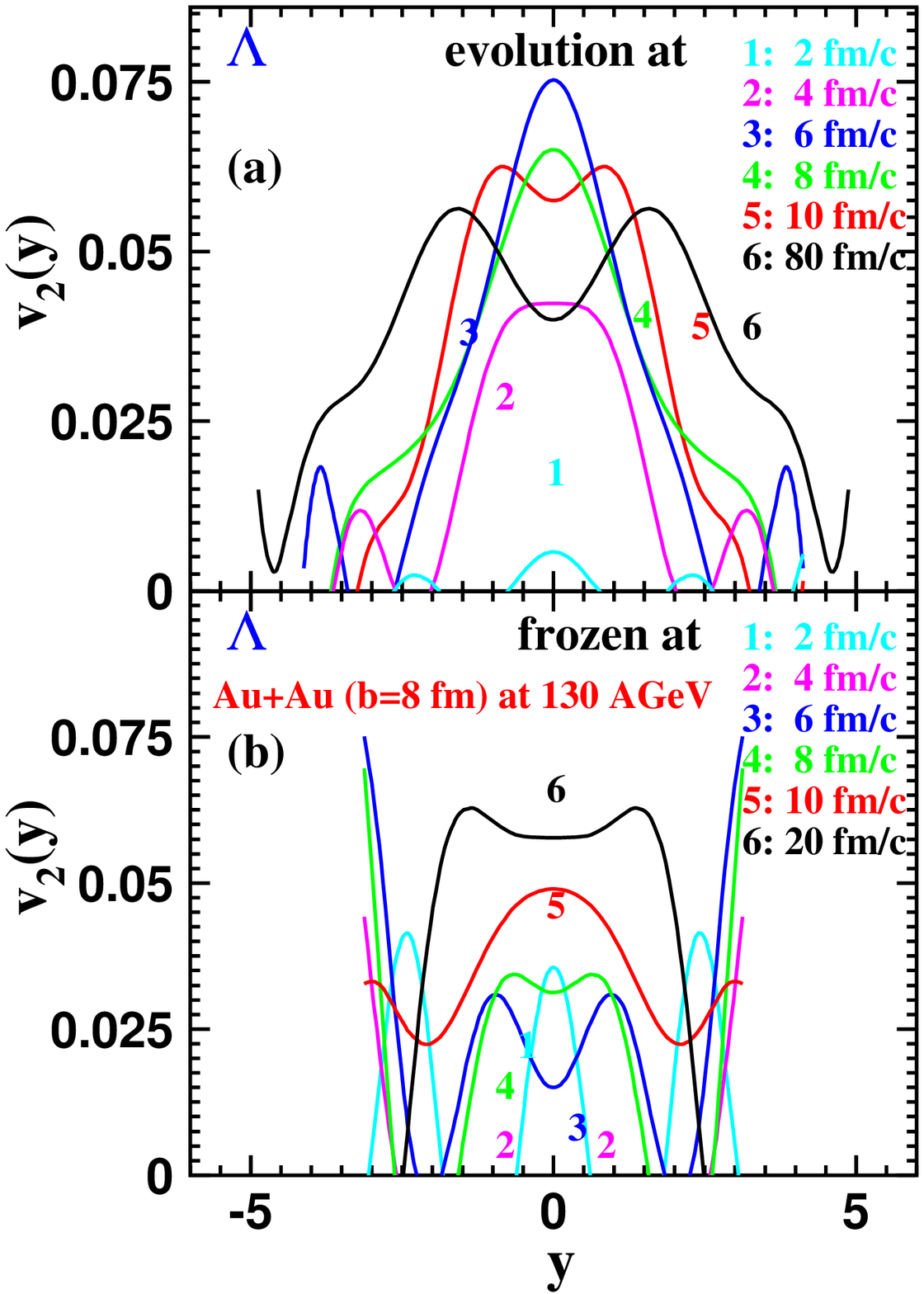}{72mm}  
\caption{
$d^2N/dp_x dp_y$ distributions for $\Lambda$ and $K$ in Au+Au
collisions with $b = 8$ fm at $\sqrt{s} = 130$ AGeV.
}
\label{fig4}
\end{minipage}
\end{figure}

Microscopic models based on string phenomenology and transport theory 
are able to reproduce many features of the elliptic flow at
ultrarelativistic energies \cite{BlSt02, LPX99,ell_fl01}.
However, the quantitative agreement with the data is often not so
good. Particularly, magnitude of the distributions $v_2(\eta)$ or
$v_2(p_t \geq 1.5$\,GeV/$c)$ appears to be too high. Does it mean
that the effective EOS of hot and dense partonic-hadronic matter in
microscopic models is too soft? 
Then, the microscopic calculations \cite{br_sqm04} show the absence 
of sharp freeze-out of particles in relativistic heavy-ion collisions. 
What are the consequences of the continuous freeze-out for the $v_2$
of these particles? To study the development of the elliptic flow
ca. 20$\cdot 10^3$ gold-gold collisions with the impact parameter
$b = 8$\,fm were generated at $\sqrt{s} = 130$\, AGeV. According to
previous studies \cite{ell_fl01,star00} the elliptic flow of charged
particles is close to its maximum at this impact parameter, and the
multiplicity of secondaries is still quite high. The time evolutions
of the $v_2$ of kaons and lambdas as functions of rapidity are 
displayed in Fig.~\ref{fig3}(a). Here the snapshots of the $v_2$
profile are taken at certain time $t = t_i$, when all hadronic 
interactions are switched off and particles are propagated freely.
To avoid ambiguities, resonances were allowed to decay
according to their branching ratios. Surprisingly, at $t = 2$ fm/$c$
elliptic flow of kaons is weak. The flow continuously increases and
reaches its maximum value $v_2^K (y=0) \approx 6\%$ already at
$t = 8$ fm/$c$. From this time the elliptic flow does not increase
anymore.
Instead, it becomes broader and develops a two-hump structure with a
relatively weak dip at midrapidity. The flow seems to continue
development till the late stages of the system evolution. 
However, the contributions of survived particles to the resulting 
elliptic flow presented in Fig.~\ref{fig3}(b) reveal the
peculiar feature: the $v_2$ of kaons, which are frozen already at
$t=2$ fm/$c$, is the {\bf strongest} among the fractions of the flow
carried by kaons decoupled from the fireball later on. The later the
kaons are frozen, the weaker their flow. One can conclude that the 
strong elliptic anisotropy of kaons, which left the system early, is 
caused by the absorption of kaons in the squeeze-out direction.  

For lambdas the evolution picture of the $v_2(y)$, shown in  
Fig.~\ref{fig4}(a), is similar to that for kaons. The flow is quite
weak at $t = 2$\,fm/$c$, then it increases and gets a full strength at
midrapidity between 8\,fm/$c$ and 10\,fm/$c$, i.e. later than the  
elliptic flow of kaons. Similarly to $v_2^K (y)$, it develops a
two-hump structure, but the humps tend to dissolve at late stages of
system evolution. In contrast to this behavior, the freeze-out
decomposition picture of $v_2^\Lambda (y)$, presented in 
Fig.~\ref{fig4}(b),
does not show monotonic tendency within first 8\,fm/$c$ of the 
reaction: The flow of $\Lambda$ frozen at 2\,fm/$c$ is identical to
that of $\Lambda$ frozen at 8\,fm/$c$, whereas lambdas decoupled from
the system between 2\,fm/$c$ and 8\,fm/$c$ almost do not contribute
to the resulting elliptic flow. Lambdas, which are decoupled after
8\,fm/$c$, have significant anisotropy in the momentum space, and the 
later the lambdas are frozen, the {\bf stronger} their elliptic flow.
This picture is similar to that obtained for the development of 
pionic and nucleonic elliptic flows \cite{erice}.

\section{Conclusions}
\label{sec4}
In summary, the features of the formation and development of 
anisotropic flow in gold-gold collisions at RHIC in the microscopic 
quark-gluon string model can be stated as follows. 
(1) The directed flow of 
all hadrons exhibits antiflow alignment within the pseudorapidity range 
$\eta \leq 2$. The signal increases as the reaction becomes more 
peripheral. At midrapidity $|\eta| \le 1$, however, the generated flow 
is quite weak and consistent with zero-flow behaviour reported by the 
STAR and PHOBOS collaborations.
(2) There is no one-to-one correspondence between
the apparent elliptic flow and the contribution to the final flow
coming from the ``survived" fraction of particles.
For instance, apparent elliptic flow of kaons at $t =2$ fm/$c$ is
weak, but kaons which are already decoupled from the system at this
moment have the strongest elliptic anisotropy caused by their 
absorption in the squeeze-out direction.
Elliptic flow of hadrons is formed not only during the first few
fm/$c$, but also during the whole evolution of the system because of
continuous freeze-out of particles. (3) The time evolutions of the
mesonic flow and baryonic flow are quite different. 
The general trend in particle flow formation in microscopic models at
ultrarelativistic energies is that the earlier mesons
are frozen, the weaker their elliptic flow. In contrast, baryons
frozen at the end of the system evolution have stronger $v_2$.
Therefore, development of particle collective flow should not be
studied independently of the freeze-out picture.


\section*{References}
\begin{thebibliography}{99}

\bibitem{VZ96} Voloshin~S and Zhang~Y 1996 \ZP C {\bf 70} 665

\bibitem{QM04} 
{\it Proc. of the Quark Matter'04\/} 2004 \JPG {\bf 30} S1

\bibitem{dir_flow} 
Bravina~L~V, Faessler Amand, Fuchs~C, Zabrodin~E~E 2000 
\PR C {\bf 61} 064902
\nonum Zabrodin~E~E, Fuchs~C, Bravina~L~V, Faessler Amand
\PR C {\bf 63} 034902

\bibitem{kflow} 
Bravina~L~V, Csernai~L~P, Faessler~A, Fuchs~C, Zabrodin~E~E  
2002 \PL B {\bf 543} 217
\nonum 
Bravina~L~V, Csernai~L~P, Faessler~A, Fuchs~C, Zabrodin~E~E  
2002 \JPG {\bf 28} 1977

\bibitem{phob_dir}
Tonjes~M~B {\it et al\/} (PHOBOS Collaboration)
2004 \JPG {\bf 30} S1243

\bibitem{MCS02}
Magas~V~K, Csernai~L~P, Strottman~D 2002 \NP A {\bf 712} 167

\bibitem{BlSt02} 
Bleicher~M and St\"ocker~H 2002 \PL B {\bf 526} 309

\bibitem{ampt}
Chen~L~W, Greco~V, Ko~C~M, Kolb~P~F 2005 \PL B {\bf 605} 95

\bibitem{LPX99} Liu~H, Panitkin~S, Xu~N
1999 \PR C {\bf 59} 348

\bibitem{ell_fl01} 
Zabrodin~E~E, Fuchs~C, Bravina~L~V, Faessler Amand  
2001 \PL B {\bf 508} 184

\bibitem{br_sqm04} 
Bravina~L~V, Tywoniuk~K, Zabrodin~E~E 
2005 \JPG {\bf 31} S989 (these proceedings)

\bibitem{star00} Ackermann~K~H {\it et al} (STAR Collaboration)
2001 \PRL {\bf 86} 402

\bibitem{erice} 
Zabrodin~E~E, Bravina~L~V, Fuchs~C, Faessler Amand  
2004 {\it Prog. Part. Nucl. Phys.\/} {\bf 53} 183

\end{thebibliography}
\end{document}